\documentclass[%
 aip,
rsi,%
 amsmath,amssymb,
 reprint,%
]{revtex4-1}

\usepackage{graphicx}
\usepackage{dcolumn}
\usepackage{bm}
\usepackage{graphicx}
\usepackage{subcaption}
\usepackage{float}

\begin{document}

\preprint{AIP/123-QED}

\title{Particle-in-Cell Modeling of Laser Thomson Scattering in Low-Density Plasmas at Elevated Laser Intensities}

\author{Andrew T. Powis}
\author{Mikhail N. Shneider}%
\affiliation{Princeton University, Princeton, New Jersey 08544, USA
}%


\date{\today}

\begin{abstract}
Incoherent Thomson scattering is a non-intrusive technique commonly used for measuring local plasma density. Within low-density, low-temperature plasma's and for sufficient laser intensity, the laser may perturb the local electron density via the ponderomotive force, causing the diagnostic to become intrusive and leading to erroneous results. A theoretical model for this effect is validated numerically via kinetic simulations of a quasi-neutral plasma using the Particle-in-Cell technique.
\end{abstract}

\maketitle

\section{Introduction}

Thomson scattering is the elastic scattering of electromagnetic radiation by a free charged particle \cite{lochte1968plasma,ovsyannikov2000plasma}. Within a plasma, scattering of light is dominated by interaction with electrons due to their comparatively lighter mass than ions. The scattering differential cross-section of light with a single electron is dependent on the angle $\phi$ between the polarization axis of the incident radiation and the direction of the scattered light element

\begin{equation}
\frac{\partial \sigma}{\partial \Omega} = \frac{e^4}{16 \pi^2 \varepsilon_0^2 m^2 c^4} \sin^2{\phi} = \frac{3}{8 \pi} \sigma_T \sin^2{\phi}
\label{eq:cross_section}
\end{equation}

Where $\sigma_T = e^4/6 \pi \varepsilon_0^2 m^2 c^4 = 6.65 \times 10^{-29} \; m^2$ is the Thomson cross-section, which does not depend on the wavelength of the incident light.

When the dimensions of the scattering volume are small in comparison to the plasma Debye length $\lambda_D = (\varepsilon_0 T_e / n_e e^2)^{1/2}$ then thermal fluctuations dominate over coherent charge interactions and Thomson scattering can be considered as incoherent. Here $T_e$ and $n_e$ are the electron temperature and density respectively. Specifically, incoherence occurs when the the parameter $\alpha = \frac{\lambda}{2 \pi \lambda_D \sin (\theta /2)} \ll 1$ where $\lambda$ is the light wavelength and $\theta$ is the angle between the observation direction and the incident light. This condition is generally satisfied for optical lasers in low pressure plasmas. 

The signal can be related to the number of electrons in the focal volume $\Delta V$ and the total light flux portending the solid angle $\Delta \Omega$ by

\begin{equation}
\Phi_s = I_L n_e \Delta V \int_{\Delta \Omega} {\frac{\partial \sigma}{\partial \Omega} d \Omega}
\label{eq:scattered_flux}
\end{equation}

Where $I_L$ is the laser beam intensity.

The most important implication here is that for a fixed laser observation angle and detection system we have $\Phi_s \propto n_e$, allowing us to measure density from a calibrated photo-detector signal.

\begin{figure}[H]
\centering
\includegraphics[width=0.9\linewidth]{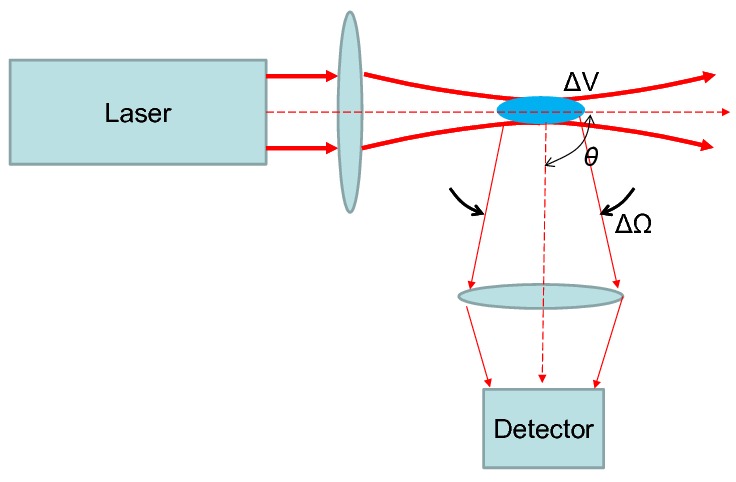}
\caption{\label{lts_setup}Typical laser Thomson scattering setup.\cite{shneider2017ponderomotive}}
\end{figure}

A typical setup for a laser Thomson scattering (LTS) system is shown in Figure \ref{lts_setup}. Measuring the given light flux at a chosen angle $\theta$ allows for measurement of the local electron density $n_e$ within the volume $\Delta V$. Broadening of the laser wavelength due to Doppler shift can also provide information on electron temperature, however here we will be concerned only with plasma density.

Laser Thomson scattering is routinely applied for measuring plasma density in a variety of equilibrium and non-equilibrium plasmas, including weakly ionized glow discharges and low-pressure discharges used in microelectronic technologies \cite{muraoka2011laser,elsabbagh2001laser,carbone2014thomson,maurmann2004thomson,crintea2009plasma}, to medium and high-pressure plasmas \cite{belostotskiy2008measurement,lee2002spectrally}, as well as within fusion plasmas \cite{peacock1969measurement,okamoto2006comparison,strickler2008thomson}.

Due to the small size of the Thomson scattering cross section LTS becomes challenging in plasma densities $n_e < 10^{11}$ $1/cm^3$. With a low number of electrons within the focal region sample volume, the scattering signal becomes weak in comparison to other sources of radiation within the plasma such as Rayleigh scattering and excitation or recombination. These issues are generally overcome by performing time averaged statistics \cite{muraoka2011laser,crintea2009plasma,elsabbagh2001laser}, with collection times ranging from minutes to hours. Time averaging, however, makes it impossible to study transient phenomena within the plasma.

An alternative approach is to increase the intensity of the laser $I_L$ such that the signal of the scattered radiation is correspondingly boosted, allowing fewer shots to be taken for the same statistical accuracy. Recently, however, it was proposed theoretically \cite{shneider2017ponderomotive} that for a low-density plasma and sufficient laser intensity $I_{max} > 10^{16}$ $W/m^2$, LTS can transition to an intrusive diagnostic, due to the effects of the ponderomotive force. The intensity of the laser beam $I_L(\mathbf{r},t)$ is non-uniform, resulting in a ponderomotive force acting on charged particles given by \cite{shneider2017ponderomotive,chen1985introduction}

\begin{equation}
\mathbf{F}_p(\mathbf{r},t) = -\frac{1}{2} \frac{q_s^2}{m_s\omega^2\varepsilon_0 c}\nabla I_L(\mathbf{r},t)
\label{eq:general_pond_force}
\end{equation}

Where $q_s$ and $m_s$ are the charge and mass respectively of plasma species $s$ and $\omega=2 \pi c / \lambda$ is the angular frequency of the laser radiation. Laser intensity profiles are generally Gaussian,

\begin{equation}
I_L(r,t) = I_0(t) \exp{(-r^2/r_L^2)}
\label{eq:intensity_profile}
\end{equation}

Where $r_L$ is the characteristic length scale of the laser intensity profile and $I_0(t)$ is a time varying maximum laser intensity, for a Gaussian pulse given by $I_0(t) = I_{max}\exp{(-(t-t_m)^2/2t_s^2)}$. In the discussion below we will also consider a Cartesian planar intensity profile, rather than cylindrical. In these instances we will let $r \rightarrow x$ for clarity.

Electrons are displaced from regions of strong laser intensity gradients by a ponderomotive ``potential", 

\begin{equation}
|U_p| \approx |\mathbf{F}(r_L,t_m)| r_L = \frac{e^2 I_{max}}{\bar{e} m_e \omega^2 \varepsilon_0 c}  
\label{eq:removal_time}
\end{equation}

Where $ \bar{e}$ is the natural number. Countering this potential is the electron pressure and ambipolar field due to charge separation of the mobile electrons from the less mobile ions.

The characteristic removal time of electrons $\tau_r$ can be estimated as,

\begin{equation}
\tau_r \approx \sqrt{\frac{2r_L m_s}{|\mathbf{F}_p |}}
\label{eq:pond_potential}
\end{equation}

For systems of interest, typically $\tau_r < 1$ $ns$, which is shorter than the laser pulse time of order $t_s \approx 10$ $ns$, and $U_p \sim T_e \sim 1$ $eV$, therefore a dip in electron density is expected to occur at the laser center. Since $\tau_r \ll t_s$, the system can be analyzed via an equilibrium force balance. Furthermore, at low pressures ($P<1$ Torr), the mean-free path of the electrons is on the order of the radius of the laser beam in the interaction region within the plasma. Therefore the effects of Joule heating and laser-induced avalanche ionization can be ignored. Within the limits of these assumptions, Reference \cite{shneider2017ponderomotive} performed a linearized analysis for a one-dimensional system, formulating an estimate for the perturbation in electron density $\delta n_e$ within the region of interaction between the laser beam and the plasma.

\begin{equation}
\frac{|\delta n_e|}{n_{0}} \approx \frac{e^2 I_0}{\bar{e} m \omega^2 \varepsilon_0 c T_e (1 + \beta r_L^2 / \lambda_D^2)}
\label{fluid-perturbation}
\end{equation}

Where $\beta = 1/2$ for cylindrical geometry and $\beta=1$ for planar geometry ($r_L \rightarrow x_L$). Therefore within the focal region of a sufficiently intense laser the local plasma density will be reduced, resulting in a weaker signal and therefore lower measured density.

In this paper, the above effect is explored via plasma simulations, demonstrating the qualitative validity of this theory.

\section{Methodology}

The plasma was simulated via the Particle-in-Cell (PIC) technique, a kinetic and self consistent approach which avoids the introduction of ad-hoc macroscopic transport models often relied on by fluid approaches. Simulations were conducted using a Princeton Plasma Physics Laboratory (PPPL) modified version of the Large-Scale Plasma code (LSP) \cite{hughes1999three}. LSP is a multi-purpose and versatile PIC code, widely benchmarked and validated within the community \cite{hughes1999three,welch2001simulation,welch2006integrated,welch2009fully,carlsson2016validation}.

PIC evolves charged and neutral macro-particles (where macro-particles are related to physical particles via a clumping parameter) in a Lagrangian sense. The motion of each macro-particles is therefore governed by Newton's second law and numerically integrated via the standard Boris algorithm \cite{boris1970relativistic}. 

The self-consistent electromagnetic field is solved in a Eulerian sense, on a grid superimposed over the simulation domain. Since the plasma is sufficiently low temperature ($T_e \sim 1$ eV), the system can be treated electrostatically, and Poisson's equation inverted to obtain the field from charge density. PPPL modifications to the code include incorporation of the latest version of the Portable Extensible Toolkit for Scientific Computation (PETSc) \cite{balay2017petsc} for improved performance and scalability of the Poisson's equation solver. In particular, Poisson's equation was inverted via PETSc's native LU factorization package.

The simulations include species present within a low-temperature pure Argon plasma, including Argon neutrals, singly charged Argon ions, and electrons. For the plasmas under consideration, the minimum ion removal time due to the ponderomotive force is greater than $20$ $\mu s$ (see Equation \ref{eq:removal_time}) and the minimum ion plasma time ($\tau_{pi}=2\pi / \omega_{pi}$ where $\omega_{pi}$ is the ion plasma frequency) is greater than $300$ $ns$. Therefore over the time scales considered within these simulations ($< 50$ $ns$) the ions and neutrals can be modeled as immobile with respect to electron dynamics. Note that if ion motion were considered it would be expected to slightly enhance the density perturbation, since the ions would move to reduce the resulting ambipolar field, thereby allowing electrons to drift further from the laser center.

Modeled collisions include coulomb collisions between all charged species and electron-neutral collisions, the latter being informed by experimental data \cite{milloy1977momentum,yamabe1983measurement}. Ionization, recombination and excitation were ignored.

Simulations were conducted in one-dimensional Cartesian geometry, with a planar laser intensity profile (Equation \ref{eq:intensity_profile} with $r\rightarrow x$), and two-dimensional Cartesian geometry, with a cylindrical laser intensity profile. Cartesian geometry avoided issues with the singularity at $r=0$ associated with cylindrical geometry. Periodic boundary conditions were used to avoid issues with sheath formation and particle loss at conducting boundaries.

The simulation domain was defined to be sufficiently large to avoid interaction between the laser and itself over the periodic domain. For one-dimensional simulations the domain was defined as $x \in [-50x_L,50x_L]$. The cell size was set as $\Delta x = x_L/20$, sufficient to resolve the plasma Debye length ($\lambda_D$) leading to a total of 2000 cells over the domain. The time step size was limited by the Courant condition for the thermal electrons, with $\Delta t = \Delta x / (10 v_{th,e})$. Resulting in a time step of around 0.83 picoseconds, and approximately $60,000$ time steps per simulation. A large number of particles per cell were required to capture the effects of the ponderomotive force over the inherent statistic noise of PIC simulations, with 10,000 electrons per cell and 10 neutrals and 10 ions per cell being deemed appropriate.

Two-dimensional simulations were conducted to quantify the discrepancy between the planar laser intensity profile used in one-dimension simulations, and a cylindrical laser profile. The two-dimensional domain was defined as $x,y \in [-5r_L,5r_L]$. With an identical cell size this led to a grid of $200 \times 200$ cells. It is shown below that this reduction in domain size has little to no effect on the results (see Figure \ref{fig:convergence_studies}). Time step and total number of steps remained the same, the number of electrons per cell was maintained at 10,000 and the number of ions and neutrals increased to 100 per cell.

The ponderomotive force due to the laser was modeled as an additional ``external" force (see Equation \ref{eq:general_pond_force} with $r \rightarrow x$). With the one-dimensional laser intensity profile (Equation \ref{eq:intensity_profile}), this led to an acceleration of each electron due to the ponderomotive force $\mathbf{a}_p$ given by

\begin{equation}
\mathbf{a}_p = \left(\frac{e^2}{m_e^2}\right) \frac{1}{\omega^2\varepsilon_0 c} \frac{x}{x_L^2} I_0(t) \exp{(-x^2/x_L^2)} \hat{\mathbf{x}}
\label{eq:electron_pond}
\end{equation}

Since the mass-to-charge ratio remains identical between real plasma particles and PIC macro-particles, Equation \ref{eq:electron_pond} demonstrates that the acceleration of a macro-particle due to the ponderomotive force will be identical to that of a real electron, a fact which is critical to maintaining self-consistency between the PIC simulations and a real plasma.

The plasma is initially neutral, with ion and electron densities of $n_0$ and neutral pressure $P$. The laser intensity is then ramped via the temporal intensity profile. Physical properties for each simulation are listed in Table \ref{tab:values}.

\begin{table}
\centering
\begin{tabular}{|c|c|c|c|}
\hline
Property & Symbol & Value(s) & Units \\\hline
Plasma Density & $n_0$ & $\{10^{8},\mathbf{10^{9}},10^{10}\}$ & $1/cm^3$\\
Electron Temperature & $T_e$ & $1$ & $eV$\\
Ion Temperature & $T_i$ & $293$ & $K$\\
Neutral Temperature & $T_n$ & $293$ & $K$\\
Neutral Pressure & $P$ & $0.1$ & $Torr$\\
Laser Wavelength & $\lambda$ & $\{\mathbf{532},1060\}$ & $nm$\\
Maximum Laser Intensity & $I_{max}$ & $\{ 1,5,\mathbf{10} \} \times 10^{16}$ & $W/m^2$\\
Laser Radius (2D) & $r_L$ & $100 $ & $\mu m$\\
Laser Width (1D) & $x_L$ & $100$ & $\mu m$\\
Laser Pulse Time & $t_s$ & $10$ & $ns$\\
Laser Pulse Center Time & $t_m$ & $25$ & $ns$\\
Total Simulation Time & $t_{end}$ & $50$ & $ns$\\\hline
\end{tabular}
\caption{\label{tab:values}Physical properties for one-dimensional simulations and two-dimensional simulation (bolded in lists). Quantities listed in braces show values explored during the parameter scan.}
\end{table}

The primary result of interest is the temporal electron density profile $n_e(x,t)$. A dip in electron density is expected at the center of the simulation domain. Therefore a primary quantity of interest will be the change in density at the laser center $\delta n(t) = (n_0 - n_e(0,t))/n_0$.

Of critical importance for experimentalists is the signal received due to laser light which is Thomson scattered by the plasma electrons. This value is then correlated to electron and therefore plasma density. If the electron density is non-uniform then the total signal strength is given by $S(t) = \int_{-L}^{L} I_L(x,t) n_e(x,t) dx$. Where $L$ is the limit of the simulation domain. An equivalently two-dimensional integral would be appropriate for two-dimensional simulations. 

Traditionally it is expected that the plasma density will remain uniform within the region of incident laser light, and therefore $n_e(x,t)=n_0$, leading to a signal $S_0(t)$. The purpose of this paper, however, is to confirm numerically that this is not always the case. Therefore another primary quantity of interest will be the change in signal strength $\delta s(t) = (S_0(t) - S(t))/S_0(t)$.

\section{Results \& Discussion}

\begin{figure}[h]
\centering
\begin{subfigure}[b]{.8\linewidth}
\includegraphics[width=\linewidth]{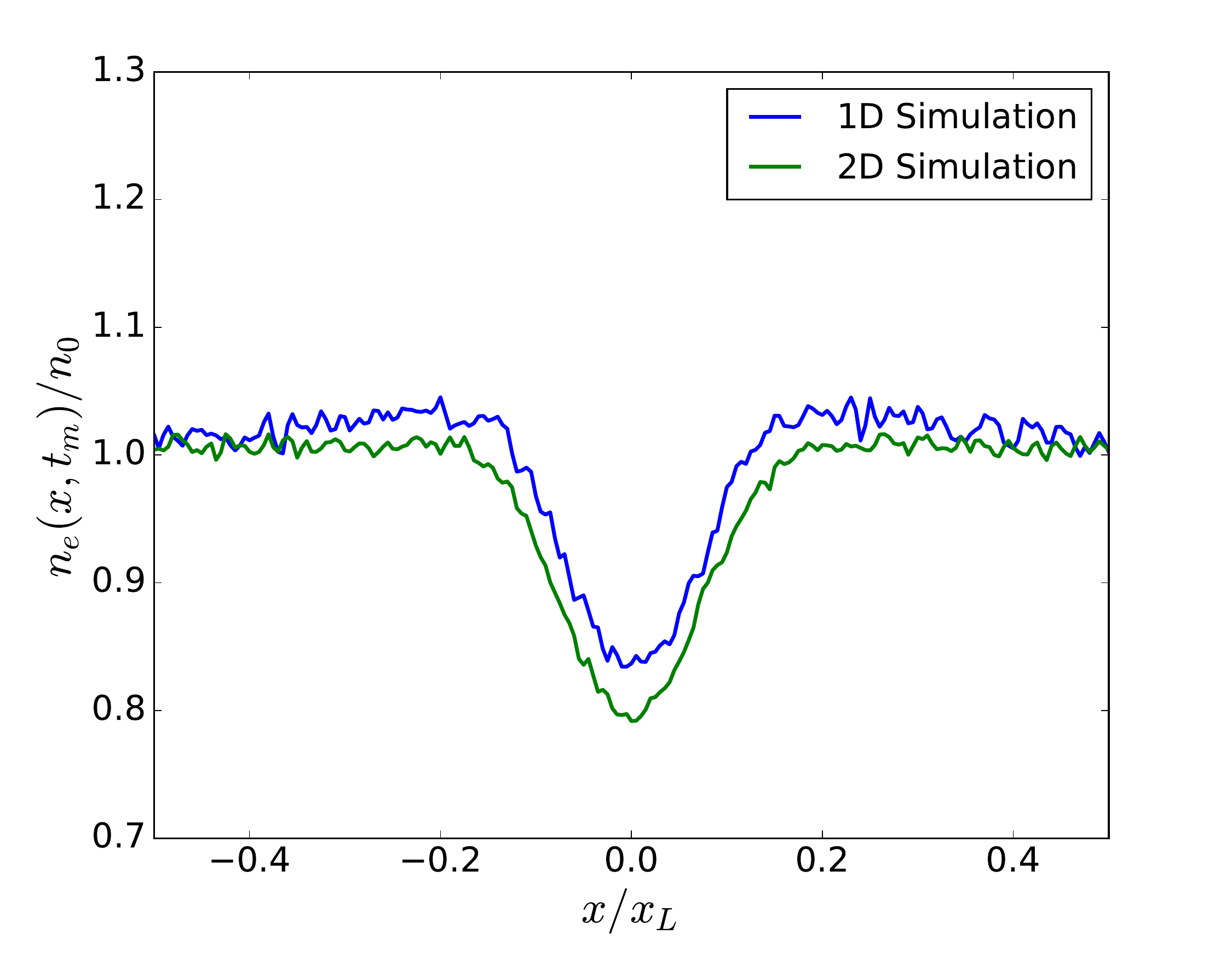}
\caption{}
\end{subfigure}

\begin{subfigure}[b]{.8\linewidth}
\includegraphics[width=\linewidth]{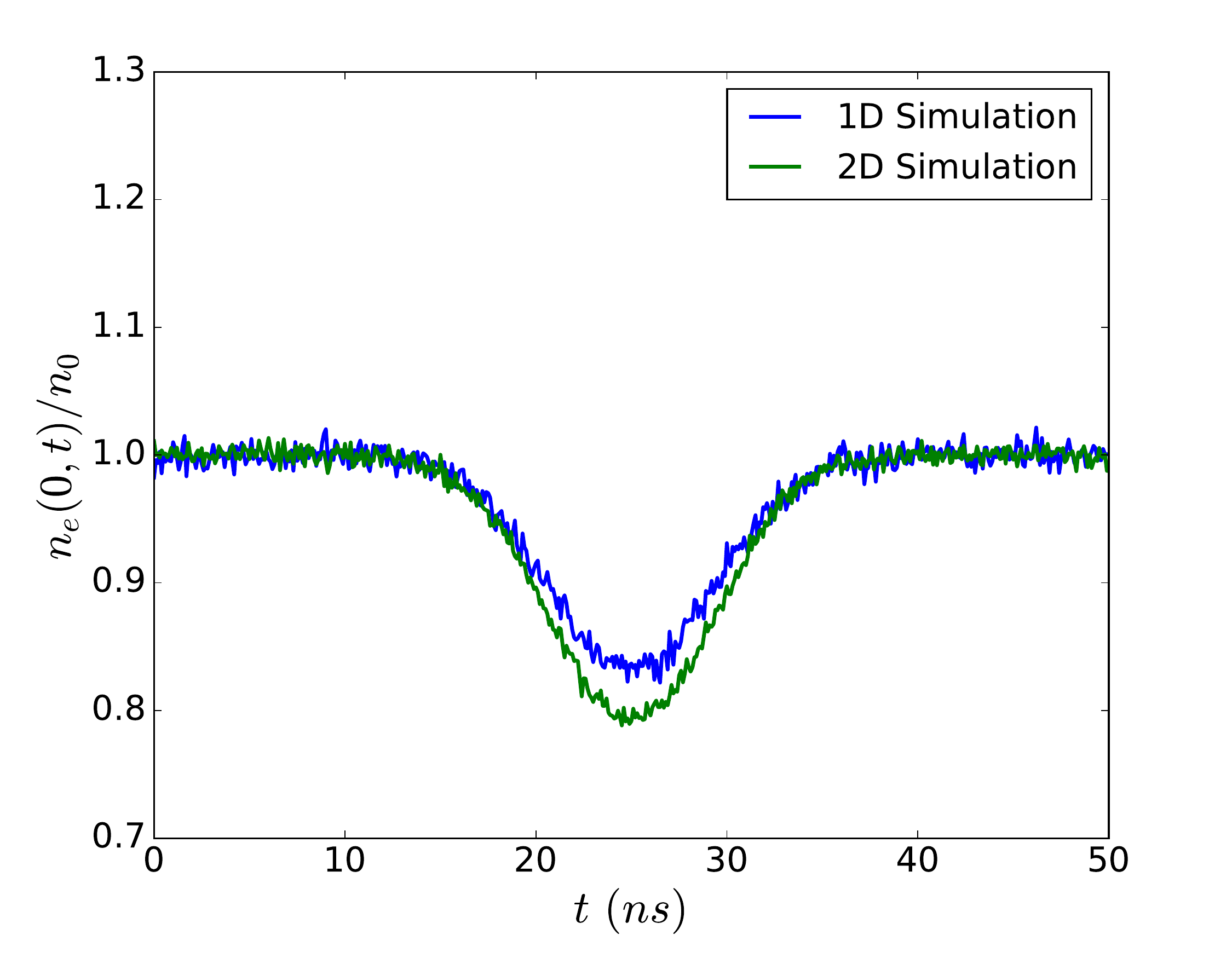}
\caption{}
\end{subfigure}

\begin{subfigure}[b]{.8\linewidth}
\includegraphics[width=\linewidth]{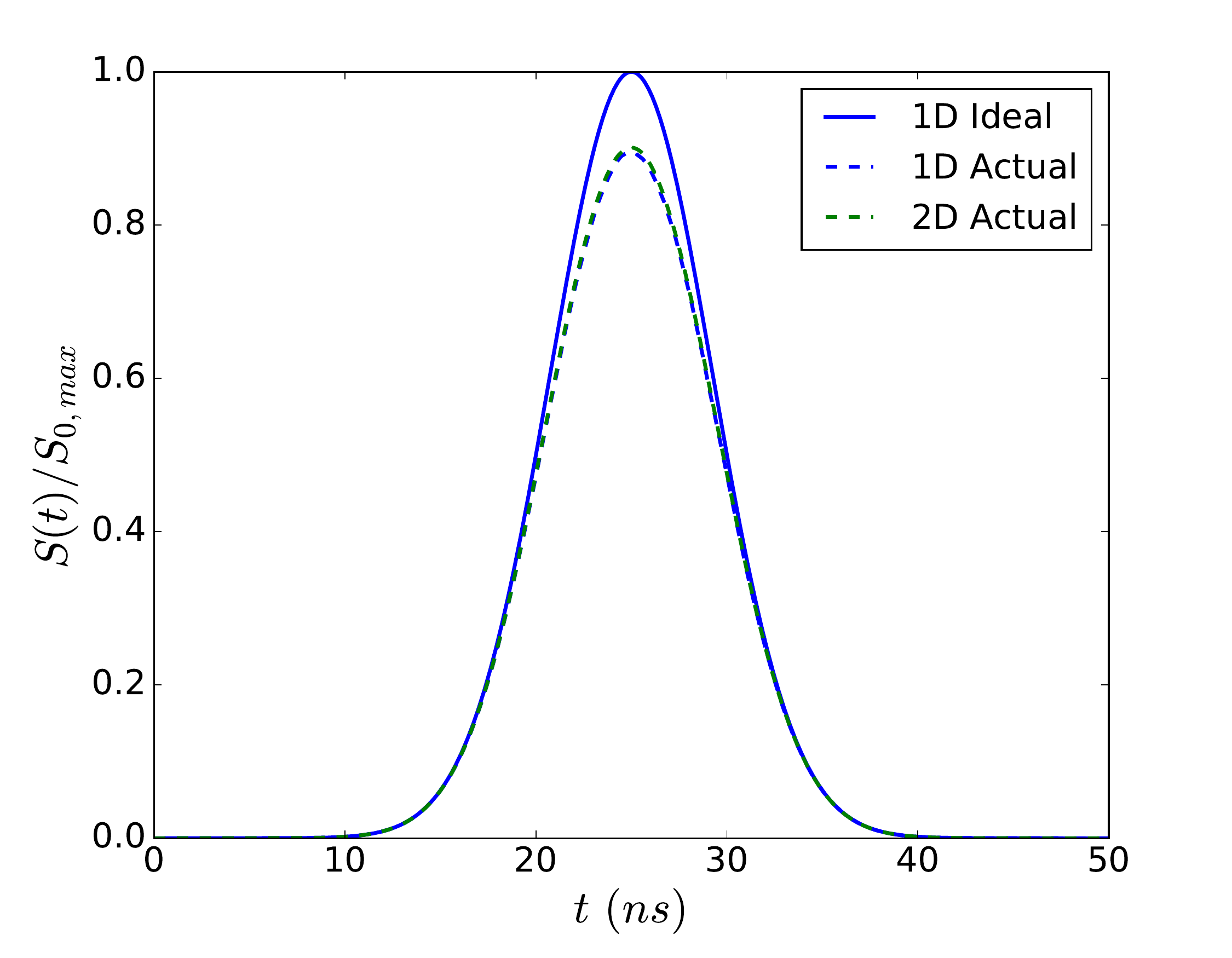}
\caption{}
\end{subfigure}

\caption{One and two-dimensional (cylindrical laser intensity) simulation results for plasma density $n_0=10^{9}$ $1/cm^3$, maximum laser intensity $I_{max}=10^{17}$ $W/m^2$ and laser wavelength $\lambda = 532$ $nm$. a) Electron plasma density distribution $n_e(x,t=t_m)$. b) Density at the center of the domain $n_e(0,t)$ against time. c) Normalized ideal signal $S_0(t)$ against actual signal $S(t)$ as would be measured by an LTS apparatus.}
\label{fig:example_result}
\end{figure}

The following results demonstrate the influence of the laser ponderomotive force for one and two-dimensional (cylindrical laser intensity profile) simulations with background plasma density of $n_0=10^{9}$ $1/cm^3$, maximum laser intensity of $I_{max}=10^{17}$ $W/m^2$ and laser wavelength of $\lambda = 532$ $nm$ (parameters highlighted in Table \ref{tab:values}).

Figure \ref{fig:example_result}a) shows the density profile at time $t=t_m$, with a dip in density at the center of the domain clearly visible. A dip in plasma density of 16.3\% and 20.8\% is observed at the center of the domain for the one and two-dimensional simulations respectively. The one-dimensional dip here compares favourably with a predicted dip of 16.5\% from Equation \ref{fluid-perturbation}, and within 5\% of the two-dimensional result. Figure \ref{fig:example_result}b) shows a plot of the density at the center of the domain $n_e(0,t)$ against simulation time. Figure \ref{fig:example_result}c) shows a plot of the ideal signal $S_0(t)$ and the actual signal $S(t)$ against simulation time, where the signal is normalized with respect to the maximum ideal signal $S_{0,max}$. A dip in signal strength of 10.5\% and 9.9\% is observed for one and two-dimensional simulations respectively, this dip could easily be mistaken for a reduced plasma density.

\begin{figure}
\centering
\includegraphics[width=0.8\linewidth]{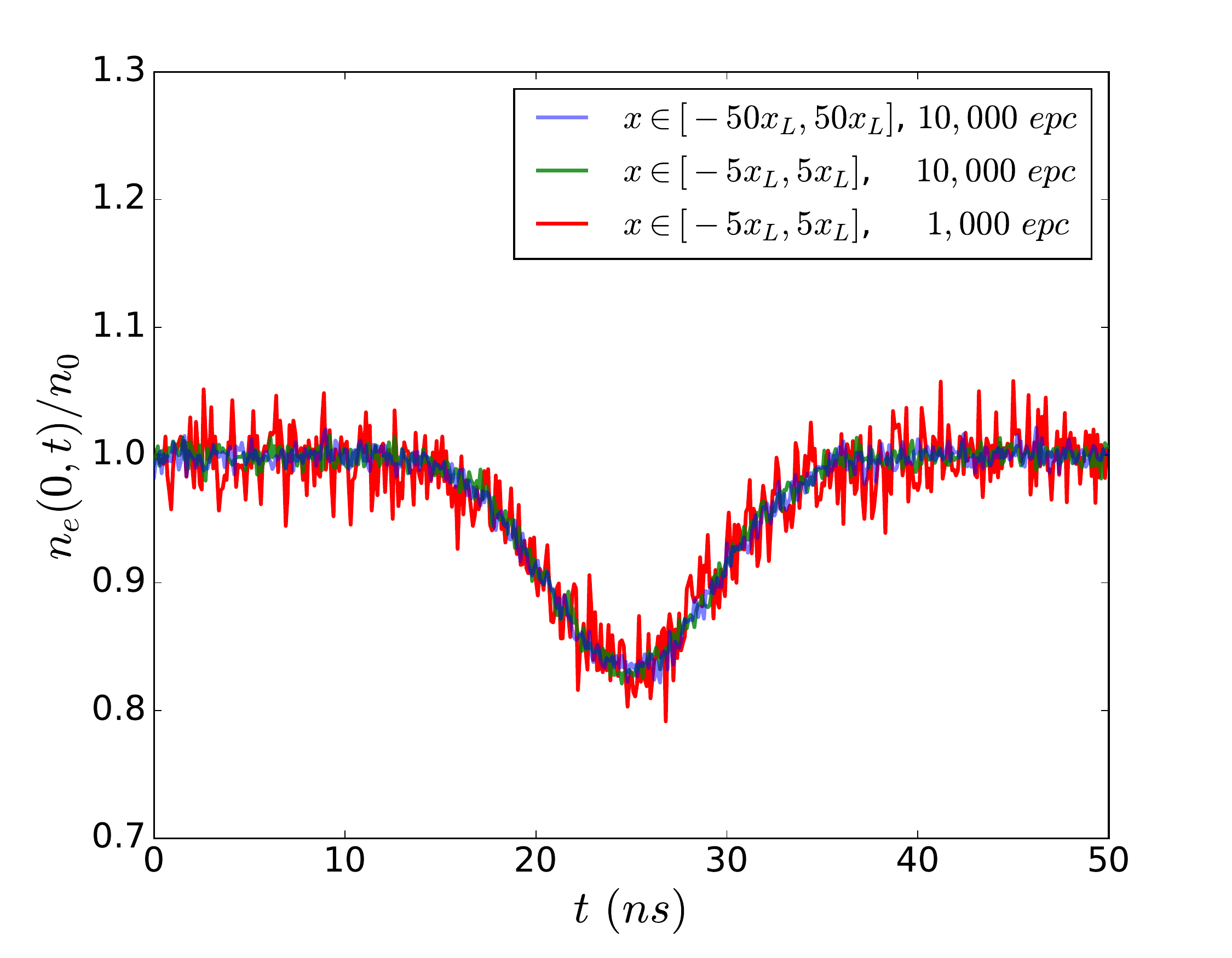}
\caption{\label{fig:convergence_studies}Plot of density at the center of the domain $n_e(0,t)$ against time for identical plasma conditions with reduced domain size (green line) and both reduced domain size and number of simulation particles (red).}
\end{figure}

Figure \ref{fig:convergence_studies} shows the same one-dimensional results from Figure \ref{fig:example_result} as well as two further simulations with identical physical parameters. The line shown in green however is for a reduced simulation domain, and the red line shows results for a simulation with reduced domain and reduced number of electrons-per-cell (epc). Together these results demonstrate numerical convergence for the choice of domain size and number of simulated particles.

\begin{figure*}
\centering
\begin{subfigure}[b]{.4\linewidth}
\includegraphics[width=\linewidth]{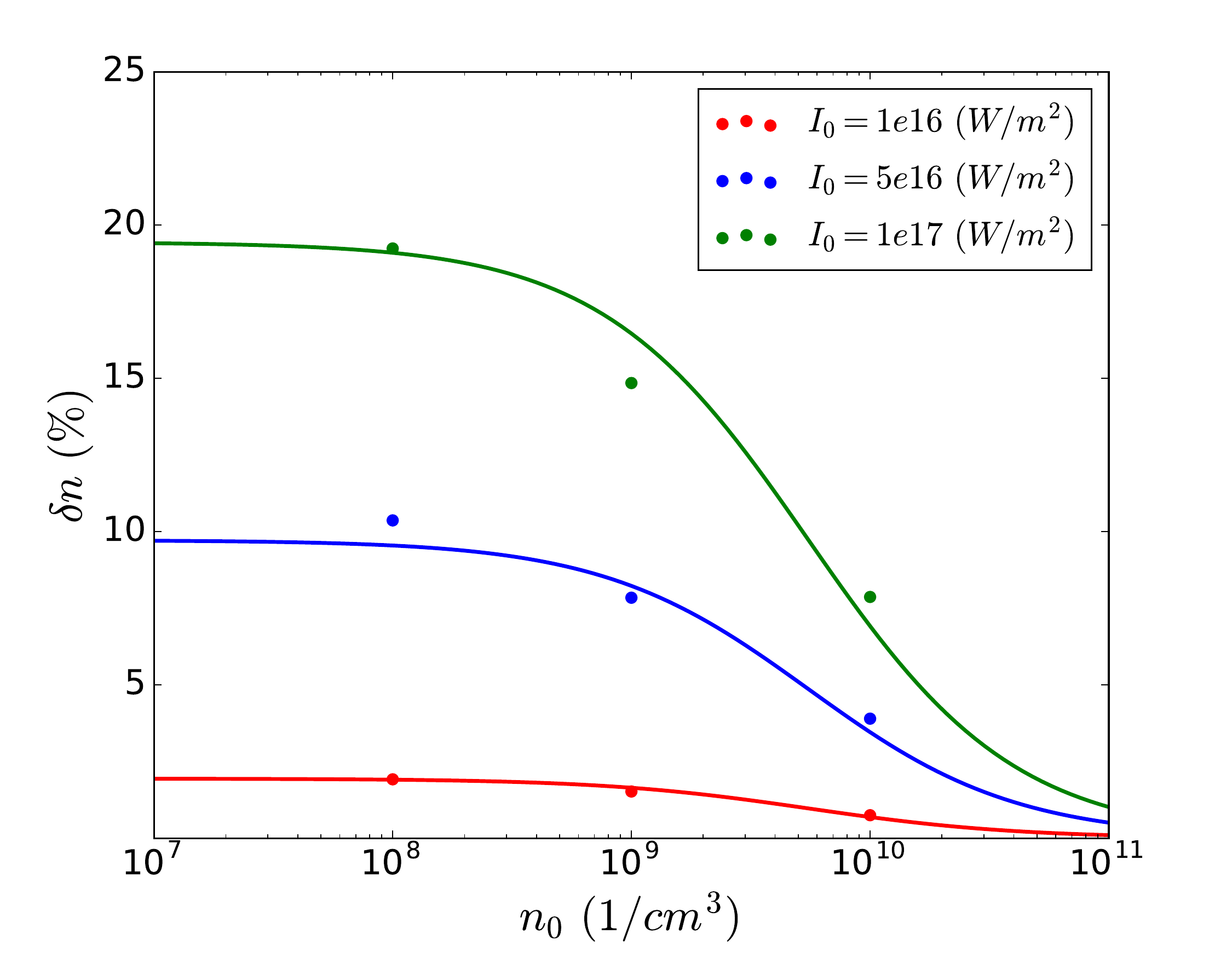}
\caption{}
\end{subfigure}
\begin{subfigure}[b]{.4\linewidth}
\includegraphics[width=\linewidth]{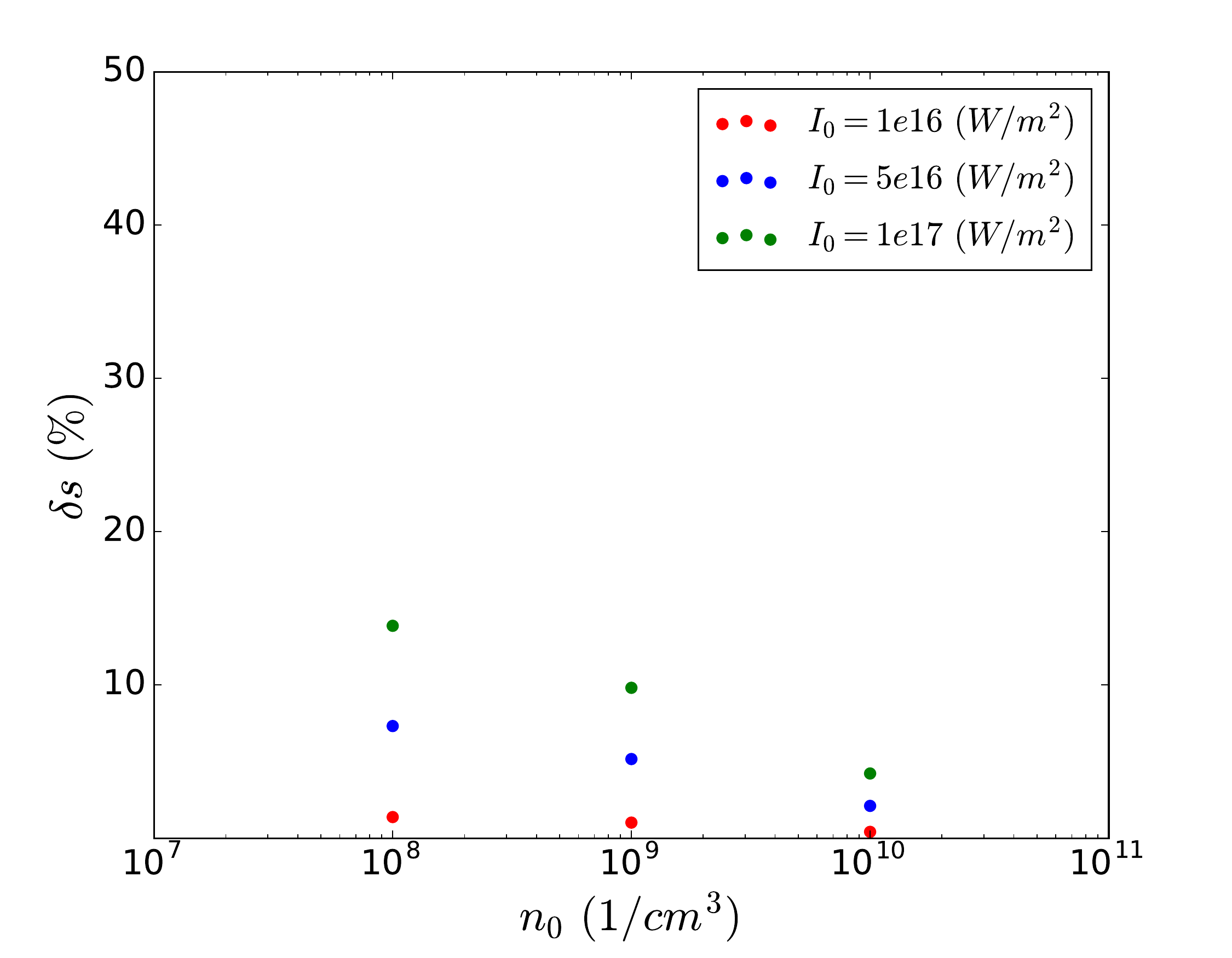}
\caption{}
\end{subfigure}

\begin{subfigure}[b]{.4\linewidth}
\includegraphics[width=\linewidth]{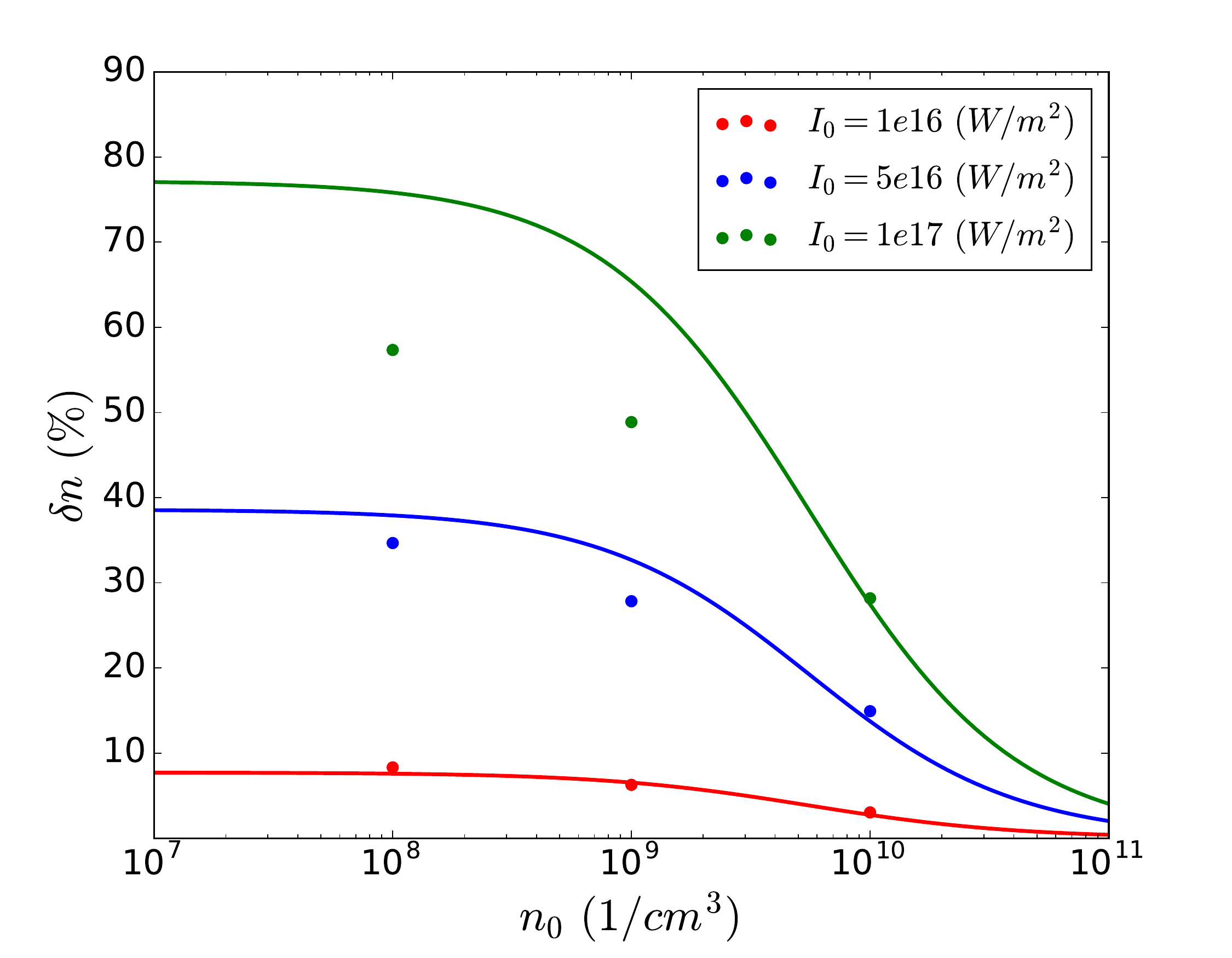}
\caption{}
\end{subfigure}
\begin{subfigure}[b]{.4\linewidth}
\includegraphics[width=\linewidth]{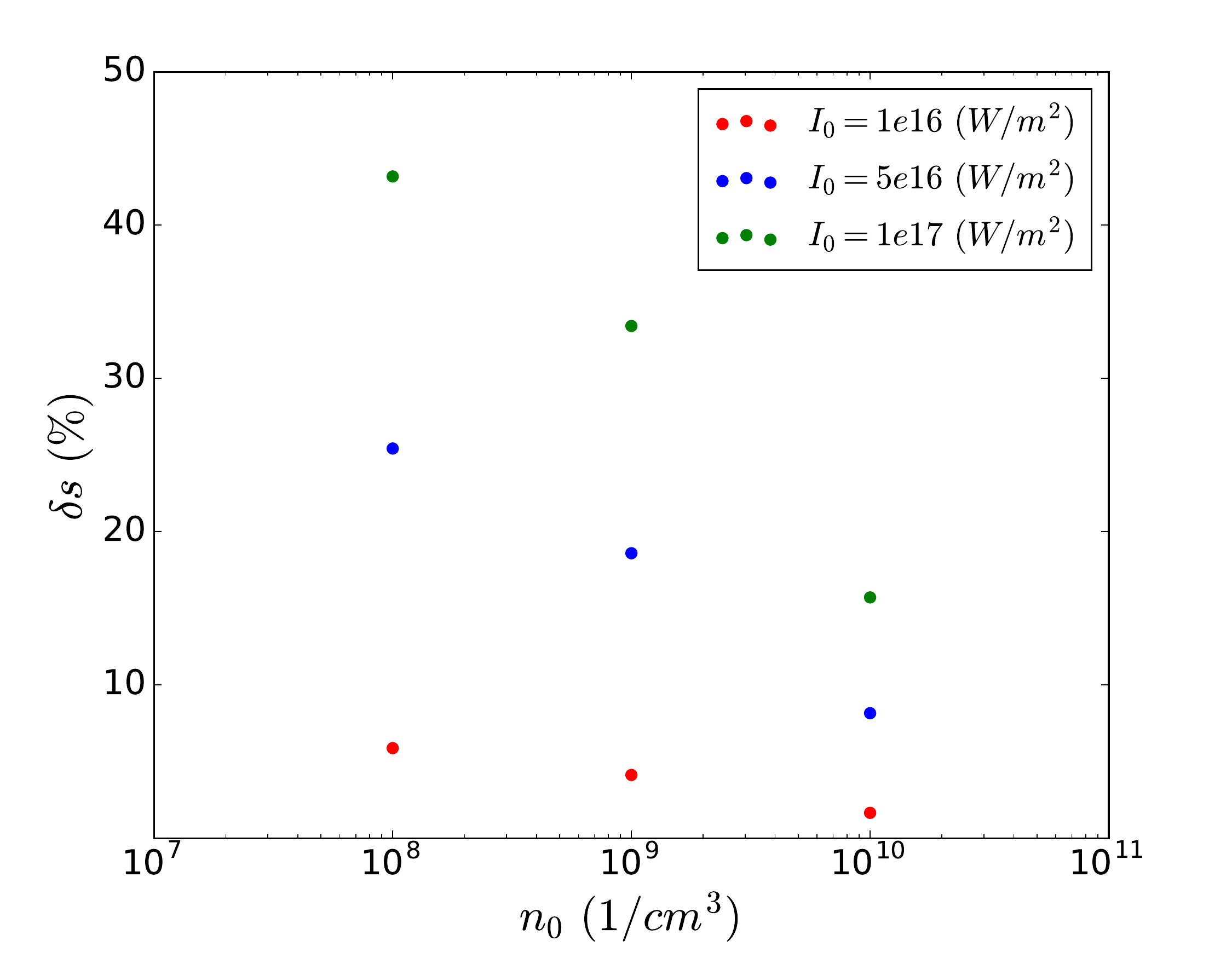}
\caption{}
\end{subfigure}

\caption{Plot of a) $\delta n$ and b) $\delta s$, for various plasma densities $n_0$ and laser intensities $I_{max}$ for laser wavelength of $532$ $nm$. Plot of c) $\delta n$ and d) $\delta s$, for laser wavelength of $1060$ $nm$. Lines on Figure's a) and c) correspond to the theoretical approximation given by Equation \ref{fluid-perturbation} with $\beta = 1$, and dots correspond to simulations.}
\label{fig:parameter_scan}
\end{figure*}

The maximum change in density $\delta n$ and maximum change in signal $\delta s$ were then calculated for each of the densities $n_0$, laser intensities $I_{max}$, and laser wavelengths $\lambda$ given in Table \ref{tab:values}. The results for $\lambda=532$ $nm$ and $\lambda=1060$ $nm$ are shown in Figure \ref{fig:parameter_scan}.

There is good agreement between simulations and the linearized theory described by Equation \ref{fluid-perturbation}. Quantitative agreement is best for low laser intensities and smaller laser wavelengths (resulting in a weaker ponderomotive force). Disagreement ensues for higher intensities and longer wavelengths as the system evolves into a non-linear state. Despite these quantitative discrepancies, the primary result is clear. In the worst case scenario, the effect of the laser on the plasma can result in a 45\% reduction in total signal strength, compared to the assumed signal strength for a uniform plasma. If the above effects were not taken into consideration, this would correspond to an equivalent 45\% error in measurement of plasma density.

\section{Conclusion}
PIC simulations were employed to simulate the effect of a diagnostic laser on a low density plasma within an incoherent LTS apparatus. It was shown that at elevated intensities the laser can evolve into an intrusive diagnostic technique, resulting in erroneous measurements of plasma density via standard experimental techniques. Experimentalists should take this into consideration when attempting to apply LTS to the measurement of low density plasmas, and where possible avoid boosting signal by increasing the diagnostic laser intensity.

\section*{Acknowledgements}
We would like to acknowledge Dr. Igor D. Kaganovich, Dr. Yevgeny Raitses, Dr. Johan Carlsson and Prof. Uwe Czarnetzki for their useful input and discussions.

This work was supported partially by the US Department of Energy, Office of Science, Fusion Energy Sciences under contract No DE-AC02-09CH11466.

\nocite{*}
\bibliography{aipsamp}

\begin{thebibliography}{23}%
\makeatletter
\providecommand \@ifxundefined [1]{%
 \@ifx{#1\undefined}
}%
\providecommand \@ifnum [1]{%
 \ifnum #1\expandafter \@firstoftwo
 \else \expandafter \@secondoftwo
 \fi
}%
\providecommand \@ifx [1]{%
 \ifx #1\expandafter \@firstoftwo
 \else \expandafter \@secondoftwo
 \fi
}%
\providecommand \natexlab [1]{#1}%
\providecommand \enquote  [1]{``#1''}%
\providecommand \bibnamefont  [1]{#1}%
\providecommand \bibfnamefont [1]{#1}%
\providecommand \citenamefont [1]{#1}%
\providecommand \href@noop [0]{\@secondoftwo}%
\providecommand \href [0]{\begingroup \@sanitize@url \@href}%
\providecommand \@href[1]{\@@startlink{#1}\@@href}%
\providecommand \@@href[1]{\endgroup#1\@@endlink}%
\providecommand \@sanitize@url [0]{\catcode `\\12\catcode `\$12\catcode
  `\&12\catcode `\#12\catcode `\^12\catcode `\_12\catcode `\%12\relax}%
\providecommand \@@startlink[1]{}%
\providecommand \@@endlink[0]{}%
\providecommand \url  [0]{\begingroup\@sanitize@url \@url }%
\providecommand \@url [1]{\endgroup\@href {#1}{\urlprefix }}%
\providecommand \urlprefix  [0]{URL }%
\providecommand \Eprint [0]{\href }%
\providecommand \doibase [0]{http://dx.doi.org/}%
\providecommand \selectlanguage [0]{\@gobble}%
\providecommand \bibinfo  [0]{\@secondoftwo}%
\providecommand \bibfield  [0]{\@secondoftwo}%
\providecommand \translation [1]{[#1]}%
\providecommand \BibitemOpen [0]{}%
\providecommand \bibitemStop [0]{}%
\providecommand \bibitemNoStop [0]{.\EOS\space}%
\providecommand \EOS [0]{\spacefactor3000\relax}%
\providecommand \BibitemShut  [1]{\csname bibitem#1\endcsname}%
\let\auto@bib@innerbib\@empty
\bibitem [{\citenamefont {Lochte-Holtegreven}(1968)}]{lochte1968plasma}%
  \BibitemOpen
  \bibfield  {author} {\bibinfo {author} {\bibfnamefont {W.}~\bibnamefont
  {Lochte-Holtegreven}},\ }\href@noop {} {\bibfield  {journal} {\bibinfo
  {journal} {Amsterdam: North-Holland Publication Co., 1968, edited by
  Lochte-Holtegreven, W.}\ } (\bibinfo {year} {1968})}\BibitemShut {NoStop}%
\bibitem [{\citenamefont {Ovsyannikov}\ and\ \citenamefont
  {Zhukov}(2000)}]{ovsyannikov2000plasma}%
  \BibitemOpen
  \bibfield  {author} {\bibinfo {author} {\bibfnamefont {A.}~\bibnamefont
  {Ovsyannikov}}\ and\ \bibinfo {author} {\bibfnamefont {M.~F.}\ \bibnamefont
  {Zhukov}},\ }\href@noop {} {\emph {\bibinfo {title} {Plasma diagnostics}}}\
  (\bibinfo  {publisher} {Cambridge Int Science Publishing},\ \bibinfo {year}
  {2000})\BibitemShut {NoStop}%
\bibitem [{\citenamefont {Shneider}(2017)}]{shneider2017ponderomotive}%
  \BibitemOpen
  \bibfield  {author} {\bibinfo {author} {\bibfnamefont {M.~N.}\ \bibnamefont
  {Shneider}},\ }\href@noop {} {\bibfield  {journal} {\bibinfo  {journal}
  {Physics of Plasmas}\ }\textbf {\bibinfo {volume} {24}},\ \bibinfo {pages}
  {100701} (\bibinfo {year} {2017})}\BibitemShut {NoStop}%
\bibitem [{\citenamefont {Muraoka}\ and\ \citenamefont
  {Kono}(2011)}]{muraoka2011laser}%
  \BibitemOpen
  \bibfield  {author} {\bibinfo {author} {\bibfnamefont {K.}~\bibnamefont
  {Muraoka}}\ and\ \bibinfo {author} {\bibfnamefont {A.}~\bibnamefont {Kono}},\
  }\href@noop {} {\bibfield  {journal} {\bibinfo  {journal} {Journal of Physics
  D: Applied Physics}\ }\textbf {\bibinfo {volume} {44}},\ \bibinfo {pages}
  {043001} (\bibinfo {year} {2011})}\BibitemShut {NoStop}%
\bibitem [{\citenamefont {ElSabbagh}\ \emph {et~al.}(2001)\citenamefont
  {ElSabbagh}, \citenamefont {Koyama}, \citenamefont {Bowden}, \citenamefont
  {Uchino},\ and\ \citenamefont {Muraoka}}]{elsabbagh2001laser}%
  \BibitemOpen
  \bibfield  {author} {\bibinfo {author} {\bibfnamefont {M.~A.~M.}\
  \bibnamefont {ElSabbagh}}, \bibinfo {author} {\bibfnamefont {H.}~\bibnamefont
  {Koyama}}, \bibinfo {author} {\bibfnamefont {M.~D.}\ \bibnamefont {Bowden}},
  \bibinfo {author} {\bibfnamefont {K.}~\bibnamefont {Uchino}}, \ and\ \bibinfo
  {author} {\bibfnamefont {K.}~\bibnamefont {Muraoka}},\ }\href@noop {}
  {\bibfield  {journal} {\bibinfo  {journal} {Japanese Journal of Applied
  Physics}\ }\textbf {\bibinfo {volume} {40}},\ \bibinfo {pages} {1465}
  (\bibinfo {year} {2001})}\BibitemShut {NoStop}%
\bibitem [{\citenamefont {Carbone}\ and\ \citenamefont
  {Nijdam}(2014)}]{carbone2014thomson}%
  \BibitemOpen
  \bibfield  {author} {\bibinfo {author} {\bibfnamefont {E.}~\bibnamefont
  {Carbone}}\ and\ \bibinfo {author} {\bibfnamefont {S.}~\bibnamefont
  {Nijdam}},\ }\href@noop {} {\bibfield  {journal} {\bibinfo  {journal} {Plasma
  Physics and Controlled Fusion}\ }\textbf {\bibinfo {volume} {57}},\ \bibinfo
  {pages} {014026} (\bibinfo {year} {2014})}\BibitemShut {NoStop}%
\bibitem [{\citenamefont {Maurmann}\ \emph {et~al.}(2004)\citenamefont
  {Maurmann}, \citenamefont {Kadetov}, \citenamefont {Khalil}, \citenamefont
  {Kunze},\ and\ \citenamefont {Czarnetzki}}]{maurmann2004thomson}%
  \BibitemOpen
  \bibfield  {author} {\bibinfo {author} {\bibfnamefont {S.}~\bibnamefont
  {Maurmann}}, \bibinfo {author} {\bibfnamefont {V.}~\bibnamefont {Kadetov}},
  \bibinfo {author} {\bibfnamefont {A.}~\bibnamefont {Khalil}}, \bibinfo
  {author} {\bibfnamefont {H.}~\bibnamefont {Kunze}}, \ and\ \bibinfo {author}
  {\bibfnamefont {U.}~\bibnamefont {Czarnetzki}},\ }\href@noop {} {\bibfield
  {journal} {\bibinfo  {journal} {Journal of Physics D: Applied Physics}\
  }\textbf {\bibinfo {volume} {37}},\ \bibinfo {pages} {2677} (\bibinfo {year}
  {2004})}\BibitemShut {NoStop}%
\bibitem [{\citenamefont {Crintea}\ \emph {et~al.}(2009)\citenamefont
  {Crintea}, \citenamefont {Czarnetzki}, \citenamefont {Iordanova},
  \citenamefont {Koleva},\ and\ \citenamefont
  {Luggenh{\"o}lscher}}]{crintea2009plasma}%
  \BibitemOpen
  \bibfield  {author} {\bibinfo {author} {\bibfnamefont {D.}~\bibnamefont
  {Crintea}}, \bibinfo {author} {\bibfnamefont {U.}~\bibnamefont {Czarnetzki}},
  \bibinfo {author} {\bibfnamefont {S.}~\bibnamefont {Iordanova}}, \bibinfo
  {author} {\bibfnamefont {I.}~\bibnamefont {Koleva}}, \ and\ \bibinfo {author}
  {\bibfnamefont {D.}~\bibnamefont {Luggenh{\"o}lscher}},\ }\href@noop {}
  {\bibfield  {journal} {\bibinfo  {journal} {Journal of Physics D: Applied
  Physics}\ }\textbf {\bibinfo {volume} {42}},\ \bibinfo {pages} {045208}
  (\bibinfo {year} {2009})}\BibitemShut {NoStop}%
\bibitem [{\citenamefont {Belostotskiy}\ \emph {et~al.}(2008)\citenamefont
  {Belostotskiy}, \citenamefont {Khandelwal}, \citenamefont {Wang},
  \citenamefont {Donnelly}, \citenamefont {Economou},\ and\ \citenamefont
  {Sadeghi}}]{belostotskiy2008measurement}%
  \BibitemOpen
  \bibfield  {author} {\bibinfo {author} {\bibfnamefont {S.~G.}\ \bibnamefont
  {Belostotskiy}}, \bibinfo {author} {\bibfnamefont {R.}~\bibnamefont
  {Khandelwal}}, \bibinfo {author} {\bibfnamefont {Q.}~\bibnamefont {Wang}},
  \bibinfo {author} {\bibfnamefont {V.~M.}\ \bibnamefont {Donnelly}}, \bibinfo
  {author} {\bibfnamefont {D.~J.}\ \bibnamefont {Economou}}, \ and\ \bibinfo
  {author} {\bibfnamefont {N.}~\bibnamefont {Sadeghi}},\ }\href@noop {}
  {\bibfield  {journal} {\bibinfo  {journal} {Applied Physics Letters}\
  }\textbf {\bibinfo {volume} {92}},\ \bibinfo {pages} {221507} (\bibinfo
  {year} {2008})}\BibitemShut {NoStop}%
\bibitem [{\citenamefont {Lee}\ and\ \citenamefont
  {Lempert}(2002)}]{lee2002spectrally}%
  \BibitemOpen
  \bibfield  {author} {\bibinfo {author} {\bibfnamefont {W.}~\bibnamefont
  {Lee}}\ and\ \bibinfo {author} {\bibfnamefont {W.~R.}\ \bibnamefont
  {Lempert}},\ }\href@noop {} {\bibfield  {journal} {\bibinfo  {journal} {AIAA
  journal}\ }\textbf {\bibinfo {volume} {40}},\ \bibinfo {pages} {2504}
  (\bibinfo {year} {2002})}\BibitemShut {NoStop}%
\bibitem [{\citenamefont {Peacock}\ \emph {et~al.}(1969)\citenamefont
  {Peacock}, \citenamefont {Robinson}, \citenamefont {Forrest}, \citenamefont
  {Wilcock},\ and\ \citenamefont {Sannikov}}]{peacock1969measurement}%
  \BibitemOpen
  \bibfield  {author} {\bibinfo {author} {\bibfnamefont {N.}~\bibnamefont
  {Peacock}}, \bibinfo {author} {\bibfnamefont {D.}~\bibnamefont {Robinson}},
  \bibinfo {author} {\bibfnamefont {M.}~\bibnamefont {Forrest}}, \bibinfo
  {author} {\bibfnamefont {P.}~\bibnamefont {Wilcock}}, \ and\ \bibinfo
  {author} {\bibfnamefont {V.}~\bibnamefont {Sannikov}},\ }\href@noop {}
  {\bibfield  {journal} {\bibinfo  {journal} {Nature}\ }\textbf {\bibinfo
  {volume} {224}},\ \bibinfo {pages} {488} (\bibinfo {year}
  {1969})}\BibitemShut {NoStop}%
\bibitem [{\citenamefont {Okamoto}\ \emph {et~al.}(2006)\citenamefont
  {Okamoto}, \citenamefont {Kado}, \citenamefont {Iida},\ and\ \citenamefont
  {Tanaka}}]{okamoto2006comparison}%
  \BibitemOpen
  \bibfield  {author} {\bibinfo {author} {\bibfnamefont {A.}~\bibnamefont
  {Okamoto}}, \bibinfo {author} {\bibfnamefont {S.}~\bibnamefont {Kado}},
  \bibinfo {author} {\bibfnamefont {Y.}~\bibnamefont {Iida}}, \ and\ \bibinfo
  {author} {\bibfnamefont {S.}~\bibnamefont {Tanaka}},\ }\href@noop {}
  {\bibfield  {journal} {\bibinfo  {journal} {Contributions to Plasma Physics}\
  }\textbf {\bibinfo {volume} {46}},\ \bibinfo {pages} {416} (\bibinfo {year}
  {2006})}\BibitemShut {NoStop}%
\bibitem [{\citenamefont {Strickler}\ \emph {et~al.}(2008)\citenamefont
  {Strickler}, \citenamefont {Majeski}, \citenamefont {Kaita},\ and\
  \citenamefont {LeBlanc}}]{strickler2008thomson}%
  \BibitemOpen
  \bibfield  {author} {\bibinfo {author} {\bibfnamefont {T.}~\bibnamefont
  {Strickler}}, \bibinfo {author} {\bibfnamefont {R.}~\bibnamefont {Majeski}},
  \bibinfo {author} {\bibfnamefont {R.}~\bibnamefont {Kaita}}, \ and\ \bibinfo
  {author} {\bibfnamefont {B.}~\bibnamefont {LeBlanc}},\ }\href@noop {}
  {\bibfield  {journal} {\bibinfo  {journal} {Review of Scientific
  Instruments}\ }\textbf {\bibinfo {volume} {79}},\ \bibinfo {pages} {10E738}
  (\bibinfo {year} {2008})}\BibitemShut {NoStop}%
\bibitem [{\citenamefont {Chen}\ and\ \citenamefont {von
  Goeler}(1985)}]{chen1985introduction}%
  \BibitemOpen
  \bibfield  {author} {\bibinfo {author} {\bibfnamefont {F.~F.}\ \bibnamefont
  {Chen}}\ and\ \bibinfo {author} {\bibfnamefont {S.~E.}\ \bibnamefont {von
  Goeler}},\ }\href@noop {} {\bibfield  {journal} {\bibinfo  {journal} {Physics
  Today}\ }\textbf {\bibinfo {volume} {38}},\ \bibinfo {pages} {87} (\bibinfo
  {year} {1985})}\BibitemShut {NoStop}%
\bibitem [{\citenamefont {Hughes}, \citenamefont {Clark},\ and\ \citenamefont
  {Simon}(1999)}]{hughes1999three}%
  \BibitemOpen
  \bibfield  {author} {\bibinfo {author} {\bibfnamefont {T.~P.}\ \bibnamefont
  {Hughes}}, \bibinfo {author} {\bibfnamefont {R.~E.}\ \bibnamefont {Clark}}, \
  and\ \bibinfo {author} {\bibfnamefont {S.~Y.}\ \bibnamefont {Simon}},\
  }\href@noop {} {\bibfield  {journal} {\bibinfo  {journal} {Physical Review
  Special Topics-Accelerators and Beams}\ }\textbf {\bibinfo {volume} {2}},\
  \bibinfo {pages} {110401} (\bibinfo {year} {1999})}\BibitemShut {NoStop}%
\bibitem [{\citenamefont {Welch}\ \emph {et~al.}(2001)\citenamefont {Welch},
  \citenamefont {Rose}, \citenamefont {Oliver},\ and\ \citenamefont
  {Clark}}]{welch2001simulation}%
  \BibitemOpen
  \bibfield  {author} {\bibinfo {author} {\bibfnamefont {D.~R.}\ \bibnamefont
  {Welch}}, \bibinfo {author} {\bibfnamefont {D.}~\bibnamefont {Rose}},
  \bibinfo {author} {\bibfnamefont {B.}~\bibnamefont {Oliver}}, \ and\ \bibinfo
  {author} {\bibfnamefont {R.}~\bibnamefont {Clark}},\ }\href@noop {}
  {\bibfield  {journal} {\bibinfo  {journal} {Nuclear Instruments and Methods
  in Physics Research Section A: Accelerators, Spectrometers, Detectors and
  Associated Equipment}\ }\textbf {\bibinfo {volume} {464}},\ \bibinfo {pages}
  {134} (\bibinfo {year} {2001})}\BibitemShut {NoStop}%
\bibitem [{\citenamefont {Welch}\ \emph {et~al.}(2006)\citenamefont {Welch},
  \citenamefont {Rose}, \citenamefont {Cuneo}, \citenamefont {Campbell},\ and\
  \citenamefont {Mehlhorn}}]{welch2006integrated}%
  \BibitemOpen
  \bibfield  {author} {\bibinfo {author} {\bibfnamefont {D.}~\bibnamefont
  {Welch}}, \bibinfo {author} {\bibfnamefont {D.}~\bibnamefont {Rose}},
  \bibinfo {author} {\bibfnamefont {M.}~\bibnamefont {Cuneo}}, \bibinfo
  {author} {\bibfnamefont {R.}~\bibnamefont {Campbell}}, \ and\ \bibinfo
  {author} {\bibfnamefont {T.}~\bibnamefont {Mehlhorn}},\ }\href@noop {}
  {\bibfield  {journal} {\bibinfo  {journal} {Physics of Plasmas}\ }\textbf
  {\bibinfo {volume} {13}},\ \bibinfo {pages} {063105} (\bibinfo {year}
  {2006})}\BibitemShut {NoStop}%
\bibitem [{\citenamefont {Welch}\ \emph {et~al.}(2009)\citenamefont {Welch},
  \citenamefont {Rose}, \citenamefont {Clark}, \citenamefont {Mostrom},
  \citenamefont {Stygar},\ and\ \citenamefont {Leeper}}]{welch2009fully}%
  \BibitemOpen
  \bibfield  {author} {\bibinfo {author} {\bibfnamefont {D.}~\bibnamefont
  {Welch}}, \bibinfo {author} {\bibfnamefont {D.}~\bibnamefont {Rose}},
  \bibinfo {author} {\bibfnamefont {R.}~\bibnamefont {Clark}}, \bibinfo
  {author} {\bibfnamefont {C.}~\bibnamefont {Mostrom}}, \bibinfo {author}
  {\bibfnamefont {W.}~\bibnamefont {Stygar}}, \ and\ \bibinfo {author}
  {\bibfnamefont {R.}~\bibnamefont {Leeper}},\ }\href@noop {} {\bibfield
  {journal} {\bibinfo  {journal} {Physical review letters}\ }\textbf {\bibinfo
  {volume} {103}},\ \bibinfo {pages} {255002} (\bibinfo {year}
  {2009})}\BibitemShut {NoStop}%
\bibitem [{\citenamefont {Carlsson}\ \emph {et~al.}(2016)\citenamefont
  {Carlsson}, \citenamefont {Khrabrov}, \citenamefont {Kaganovich},
  \citenamefont {Sommerer},\ and\ \citenamefont
  {Keating}}]{carlsson2016validation}%
  \BibitemOpen
  \bibfield  {author} {\bibinfo {author} {\bibfnamefont {J.}~\bibnamefont
  {Carlsson}}, \bibinfo {author} {\bibfnamefont {A.}~\bibnamefont {Khrabrov}},
  \bibinfo {author} {\bibfnamefont {I.}~\bibnamefont {Kaganovich}}, \bibinfo
  {author} {\bibfnamefont {T.}~\bibnamefont {Sommerer}}, \ and\ \bibinfo
  {author} {\bibfnamefont {D.}~\bibnamefont {Keating}},\ }\href@noop {}
  {\bibfield  {journal} {\bibinfo  {journal} {Plasma Sources Science and
  Technology}\ }\textbf {\bibinfo {volume} {26}},\ \bibinfo {pages} {014003}
  (\bibinfo {year} {2016})}\BibitemShut {NoStop}%
\bibitem [{\citenamefont {Boris}(1970)}]{boris1970relativistic}%
  \BibitemOpen
  \bibfield  {author} {\bibinfo {author} {\bibfnamefont {J.~P.}\ \bibnamefont
  {Boris}},\ }in\ \href@noop {} {\emph {\bibinfo {booktitle} {Proc. Fourth
  Conf. Num. Sim. Plasmas, Naval Res. Lab, Wash. DC}}}\ (\bibinfo {year}
  {1970})\ pp.\ \bibinfo {pages} {3--67}\BibitemShut {NoStop}%
\bibitem [{\citenamefont {Balay}\ \emph {et~al.}(2017)\citenamefont {Balay},
  \citenamefont {Abhyankar}, \citenamefont {Adams}, \citenamefont {Brown},
  \citenamefont {Brune}, \citenamefont {Buschelman}, \citenamefont {Dalcin},
  \citenamefont {Eijkhout}, \citenamefont {Gropp}, \citenamefont {Kaushik}
  \emph {et~al.}}]{balay2017petsc}%
  \BibitemOpen
  \bibfield  {author} {\bibinfo {author} {\bibfnamefont {S.}~\bibnamefont
  {Balay}}, \bibinfo {author} {\bibfnamefont {S.}~\bibnamefont {Abhyankar}},
  \bibinfo {author} {\bibfnamefont {M.}~\bibnamefont {Adams}}, \bibinfo
  {author} {\bibfnamefont {J.}~\bibnamefont {Brown}}, \bibinfo {author}
  {\bibfnamefont {P.}~\bibnamefont {Brune}}, \bibinfo {author} {\bibfnamefont
  {K.}~\bibnamefont {Buschelman}}, \bibinfo {author} {\bibfnamefont
  {L.}~\bibnamefont {Dalcin}}, \bibinfo {author} {\bibfnamefont
  {V.}~\bibnamefont {Eijkhout}}, \bibinfo {author} {\bibfnamefont
  {W.}~\bibnamefont {Gropp}}, \bibinfo {author} {\bibfnamefont
  {D.}~\bibnamefont {Kaushik}},  \emph {et~al.},\ }\href@noop {} {\enquote
  {\bibinfo {title} {Petsc users manual revision 3.8},}\ }\bibinfo {type}
  {Tech. Rep.}\ (\bibinfo  {institution} {Argonne National Lab.(ANL), Argonne,
  IL (United States)},\ \bibinfo {year} {2017})\BibitemShut {NoStop}%
\bibitem [{\citenamefont {Milloy}\ \emph {et~al.}(1977)\citenamefont {Milloy},
  \citenamefont {Crompton}, \citenamefont {Rees},\ and\ \citenamefont
  {Robertson}}]{milloy1977momentum}%
  \BibitemOpen
  \bibfield  {author} {\bibinfo {author} {\bibfnamefont {H.}~\bibnamefont
  {Milloy}}, \bibinfo {author} {\bibfnamefont {R.}~\bibnamefont {Crompton}},
  \bibinfo {author} {\bibfnamefont {J.}~\bibnamefont {Rees}}, \ and\ \bibinfo
  {author} {\bibfnamefont {A.}~\bibnamefont {Robertson}},\ }\href@noop {}
  {\bibfield  {journal} {\bibinfo  {journal} {Australian Journal of Physics}\
  }\textbf {\bibinfo {volume} {30}},\ \bibinfo {pages} {61} (\bibinfo {year}
  {1977})}\BibitemShut {NoStop}%
\bibitem [{\citenamefont {Yamabe}, \citenamefont {Buckman},\ and\ \citenamefont
  {Phelps}(1983)}]{yamabe1983measurement}%
  \BibitemOpen
  \bibfield  {author} {\bibinfo {author} {\bibfnamefont {C.}~\bibnamefont
  {Yamabe}}, \bibinfo {author} {\bibfnamefont {S.}~\bibnamefont {Buckman}}, \
  and\ \bibinfo {author} {\bibfnamefont {A.}~\bibnamefont {Phelps}},\
  }\href@noop {} {\bibfield  {journal} {\bibinfo  {journal} {Physical Review
  A}\ }\textbf {\bibinfo {volume} {27}},\ \bibinfo {pages} {1345} (\bibinfo
  {year} {1983})}\BibitemShut {NoStop}%
\end{thebibliography}%

\end{document}